\documentclass[twocolumn,pre,aps,showpacs,epsfig,dvips,amsmath,amssymb]{revtex4}
\usepackage{eufrak}
\usepackage[dvips]{graphicx}
\usepackage{dcolumn}
\usepackage{bm}
\usepackage{epsfig}
\usepackage{multirow}
\usepackage{color}

\newcommand{\rv}{{\vec r}}

\newcommand{\xv}{{\vec x}}

\newcommand{\Tr}{{\rm Tr}}

\newcommand{\nh}{{\hat{n}}}

\newcommand{\eh}{{\hat{e}}}

\newcommand{\mm}[1]{{\bf #1}}
\newcommand{\ar}{{\zeta}}

\def\Qm{{{\mathbf Q}}}
\def\mm#1{ {\mathbf #1}}

\def\Im{{{\mathbf I}}}

\def\Lm{{{\mathbf \Lambda}}}

\DeclareMathAlphabet{\mathpzc}{OT1}{pzc}{m}{it}
\begin{document}
\title{Isotropic-Cholesteric Transition of a Weakly Chiral Elastomer Cylinder}
\author{Xiangjun Xing and Aparna Baskaran}
\affiliation{Department of Physics, Syracuse University, Syracuse, NY 13244}
\date{January 23, 2008}
\date{\today} 

\begin{abstract}
When a chiral isotropic elastomer is brought to low temperature cholesteric phase, the nematic degree of freedom tends to order and form a helix.  Due to the nemato-elastic coupling, this also leads to elastic deformation of the polymer network that is locally coaxial with the nematic order.  However, the helical structure of nematic order is incompatible with the energetically preferred elastic deformation.   The system is therefore frustrated and appropriate compromise has to be achieved between the nematic ordering and the elastic deformation.  For a strongly chiral elastomer whose pitch is much smaller than the system size, this problem has been studied by Pelcotivs and Meyer, as well as by Warner.  In this work, we study the isotropic-cholesteric transition in the weak chirality limit, where the pitch is comparable or much larger than system size.  We compare two possible solutions: a helical state as well as a double twist state.  We find that the double twist state very efficiently minimizes both the elastic free energy and the chiral nematic free energy.  On the other hand, the pitch of the helical state is strongly affected by the nemato-elastic coupling.  As a result this state is not efficient in minimizing the chiral nematic free energy.
\end{abstract}
\pacs{%
61.30.-v    
61.30.Cz    
61.30.Vx    
}

\maketitle

\section{Introduction}

\begin{figure*}[htb!]
\begin{center}
\includegraphics[height=5cm]{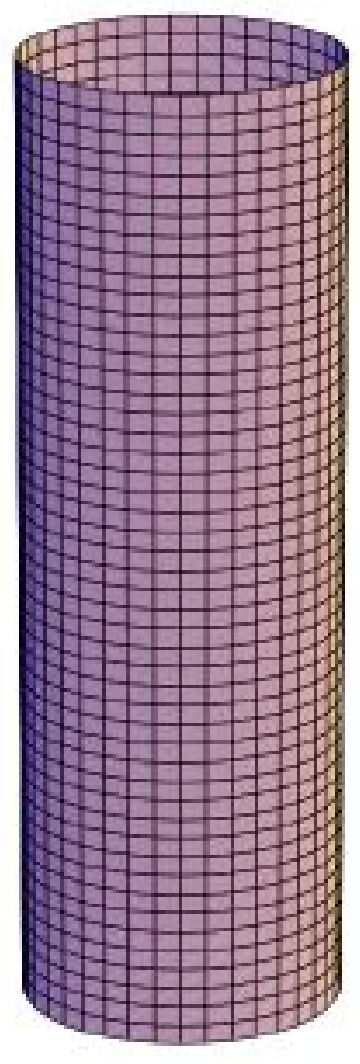}
\includegraphics[height=5cm]{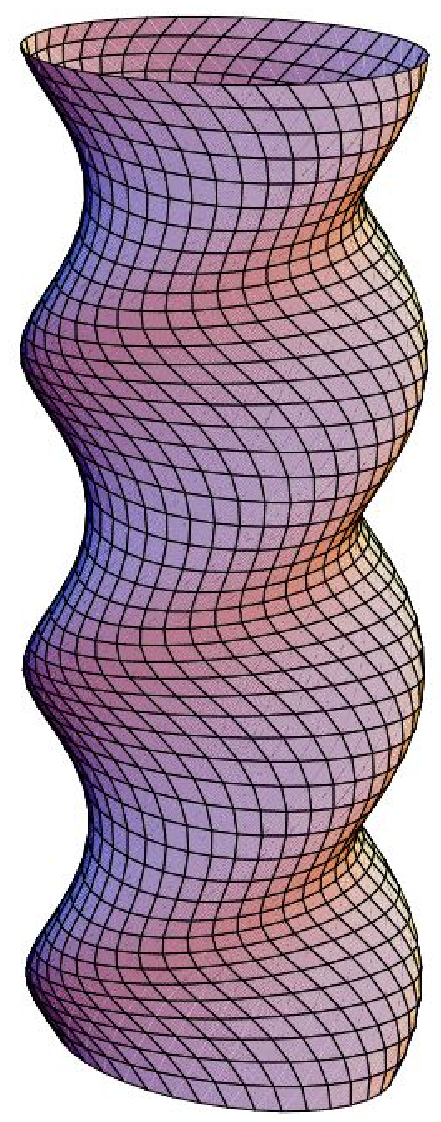}
\includegraphics[height=5cm]{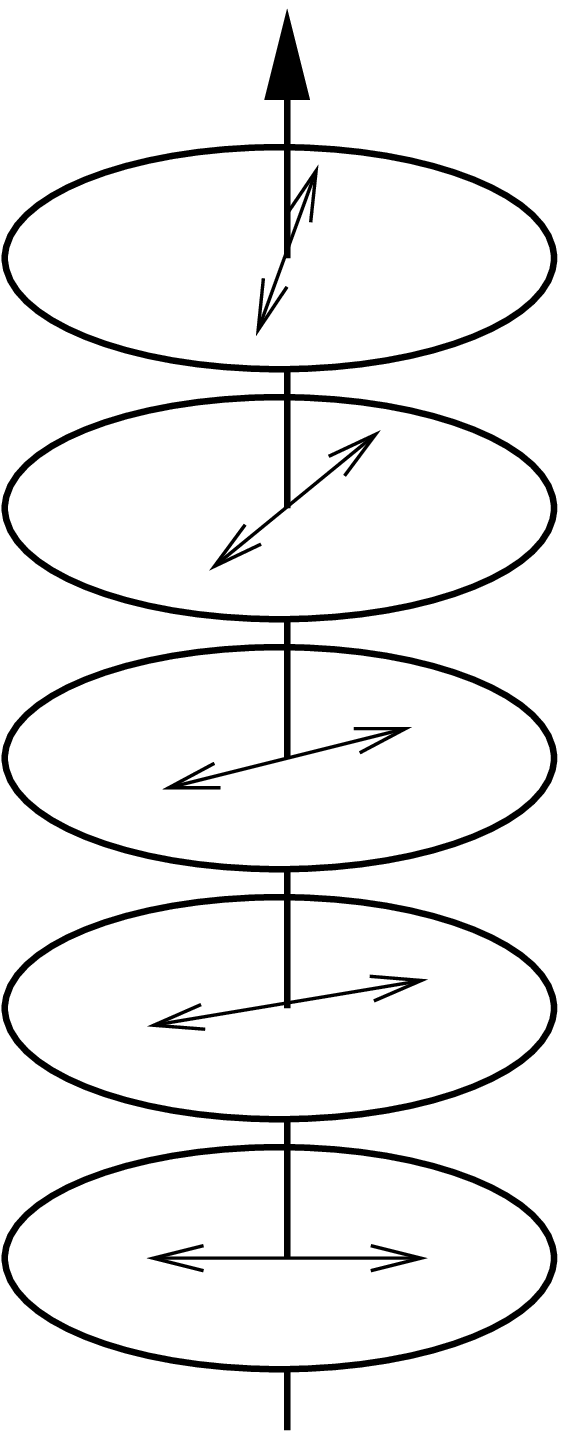}
\includegraphics[height=5cm]{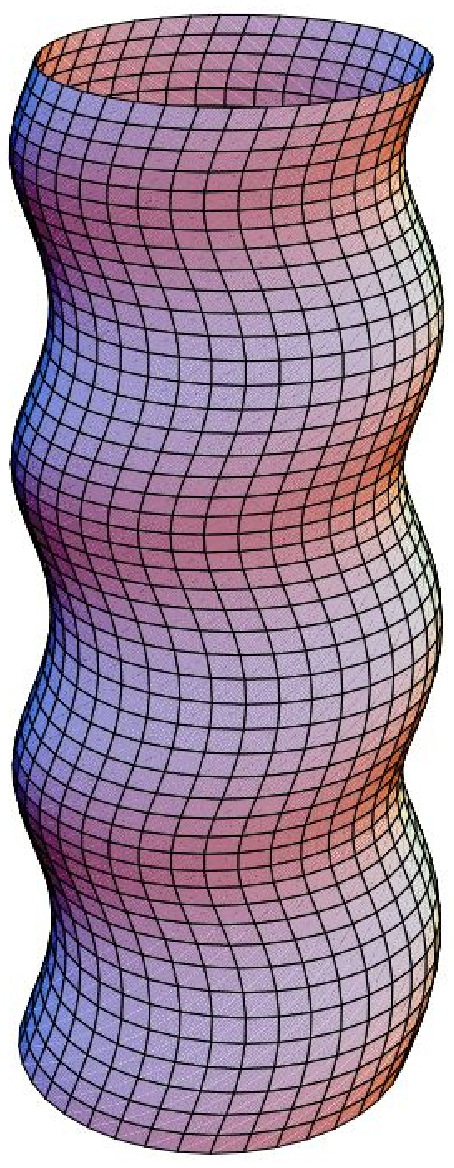}
\includegraphics[height=5cm]{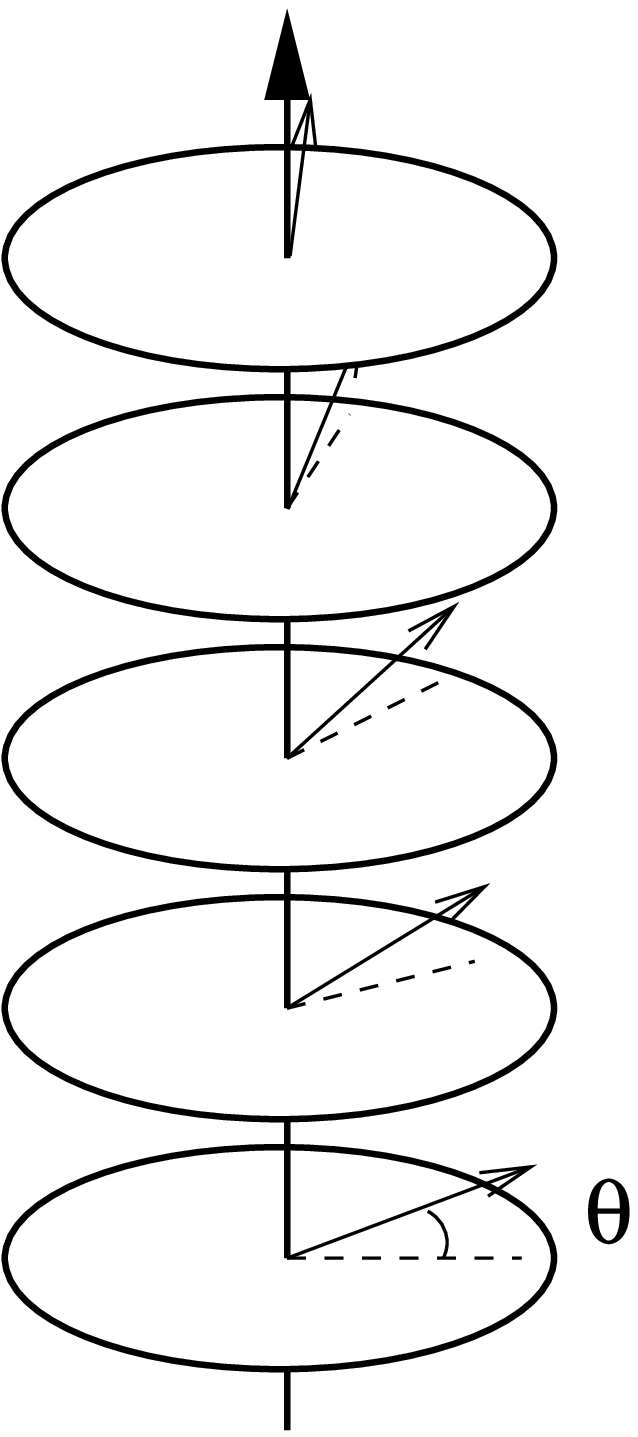}
\includegraphics[height=5cm]{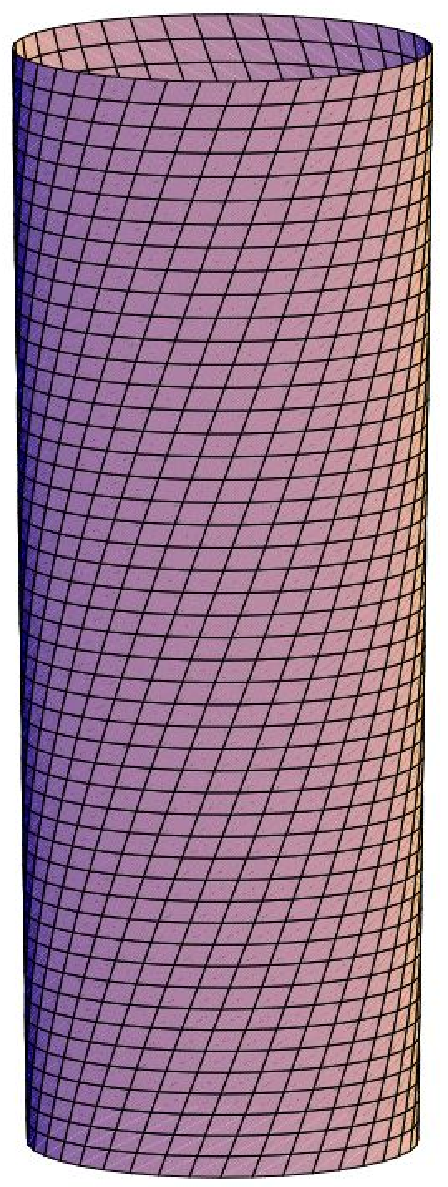}
\includegraphics[height=5cm,width=2.5cm]{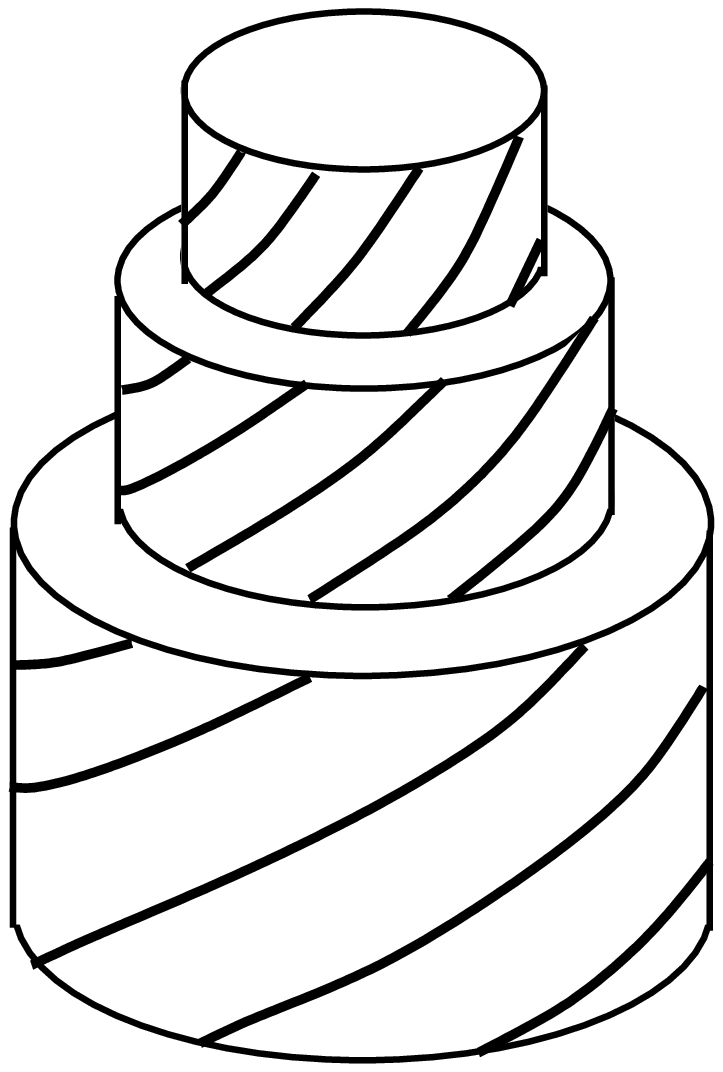}
\\
\vspace{3mm}
\includegraphics[width=15cm]{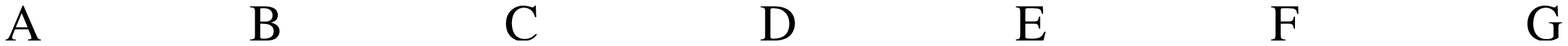}
\caption{ Elastic deformation and director pattern for various states.
A: Reference cylinder in the isotropic phase.
B: The elastic deformation of the planar helix state at weak chirality limit, studied in Sec.~\ref{Sec:helical}.
C: The nematic director pattern of the planar helix state.
D: The elastic deformation of the conical helix state, studied in references \cite{Pelcovits-Meyer-Iso-Cholesteric,Warner-iso-chiral}.
E: The director pattern of the conical helix state, $\theta$ is the conical angle.
F: The elastic deformation of the double twist state.
G: The director pattern of the double twist state. }
\label{helix}
\end{center}
\end{figure*}

An isotropic chiral elastomer can be synthesized by crosslinking a
chiral nematic polymer melt in the isotropic phase.  When such a
system is brought into the low temperature cholesteric phase, the
nematic degree of freedom orders locally and tends to form a
helical structure.   Due to the nemato-elastic coupling, the
polymer network tends to stretch along the direction of the local
nematic order, which continuously rotates along the helical axis.
For a system with cylindrical shape, this leads to strain
deformation which increase linearly with the cylinder radius, as
illustrated in Fig.~\ref{helix}B.  Its elastic energy cost is
formidably high, when the cylinder radius is much larger than the
helical pitch.  This frustration due to competition between
network elasticity and liquid crystalline ordering makes it
nontrivial to find the ground state of the system in the
cholesteric phase.

This problem was first studied by Pelcovits and Meyer
\cite{Pelcovits-Meyer-Iso-Cholesteric} using linear elasticity
theory.  In the limit of infinitely strong chirality, it is clear
that the system should first satisfy the chirality by forming a
planar helix along the cylinder axis.   On the other hand, to
avoid large strain energy, the solid can only deform uniaxially,
which implies that the nemato-elastic coupling can only be
partially satisfied.   Such a state, as illustrated in
Fig.~\ref{helix}A and \ref{helix}C, is called a planar helix state
in reference \cite{Pelcovits-Meyer-Iso-Cholesteric} and a
transverse cholesteric state in reference
\cite{Warner-iso-chiral}.   As the chirality is made weaker, a
conical helix state, where the precessing director has a
nonvanishing component along the helical axis, may constitute a
better solution.  The associated solid deformation as well as
director pattern for this conic state are illustrated in
Fig.~\ref{helix}D and \ref{helix}E respectively.   Warner
\cite{Warner-iso-chiral} carried out a nonlinear analysis of the
same problem using the neo-classical elasticity theory
\cite{LCE:WT,WarTer96}.  Nontrivial dependence of the phase
boundary on the magnitude of nematic order was identified.  The
multicritical point associated with the planar-conic transition
(where the first order transition line and the second order
transition line meet) was also analyzed.

\begin{figure}
\begin{center}
\includegraphics[height=5cm]{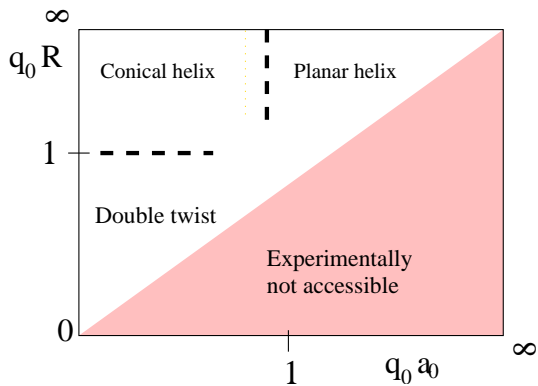}
\caption{``Phase diagram'' of a chiral cylinder in nematic phase.
The parameters $a_0$, $q_0$, and $R$ are defined in Sec.~\ref{Sec:model}.
The transition between planar helix state and conical helix state may be continuous
or discontinuous, while the transition between conical helix state and double twist state
is expected to be discontinuous.
}
\label{phase-diagram}
\end{center}
\vspace{0mm}
\end{figure}

It is implicitly assumed both in reference
\cite{Pelcovits-Meyer-Iso-Cholesteric} and reference
\cite{Warner-iso-chiral} that  the pitch of the corresponding
nematic liquid crystal system (typical $\leq 0.1\mu m$) is much
smaller than the system size, e.g. the radius of the cylinder.
That is, the elastomer is in the strong chirality limit. This is
certainly correct for many cases.  Nevertheless, the cholesteric
pitch can be continuously tuned by changing the concentration of
chiral chemical groups during polymerization.   In particular it
can be tuned to be comparable to macroscopic length scales, e.g.
the system size.   This is especially true if the system has the
shape of a thin cylinder or wire.   It is therefore interesting
and relevant to study the isotropic-cholesteric transition in the
weak chirality limit.   In this work, we carry out a nonlinear
elasticity analysis of this problem using variational methods.  We
find that in this regime, a double twist state has lower free
energy than the usual helix director pattern.  The results
obtained by Pelcovits and Meyer
\cite{Pelcovits-Meyer-Iso-Cholesteric}, Warner
\cite{Warner-iso-chiral}, as well as the authors in this work are
summarized by the ``phase diagram''  of a chiral nematic cylinder
in Fig.~\ref{phase-diagram}.

\section{Model}
\label{Sec:model}
The total free energy per unit volume of a chiral liquid crystalline elastomer crosslinked in the isotropic phase is given by
\begin{eqnarray}
f = f_{\rm el} + f_{\Qm}, \label{f-total}
\end{eqnarray}
where $f_{\rm el}$ is the neo-classical elastic free energy
\begin{eqnarray}
 f_{\rm el} = \frac{1}{2} \mu \, \Tr \,\Lm^{\rm T} \mm{l}^{-1} \Lm
 -\frac{3}{2} \,\mu,
 \label{neo-classical}
\end{eqnarray}
with
\begin{eqnarray}
\Lambda_{ia} =  \frac{\partial r_i} {\partial x_a}
\end{eqnarray}
the deformation gradient matrix defined relative to the isotropic reference state $\rv = \xv$, which is subject to the incompressibility constraint:
\begin{eqnarray}
\det \Lm \equiv 1.   \nonumber
\end{eqnarray}
As usual, the vector $\xv$ coincides with the position of the mass
points in the isotropic reference state and is referred to as the
Lagrangian coordinate. The vector $\rv$ on the other hand
describes the position of mass points in the chiral nematic
reference state (ground state that minimizes the total free
energy), and is usually referred to as the Eulerian coordinate. As
a general property of nonlinear elasticity theory, it is important
to distinguish these two coordinates properly.
The symmetric and positive definite tensor $\mm{l}$ in the neoclassical elastic free energy Eq.~(\ref{neo-classical}) is called the step length tensor \cite{WarTer96} of the current state, or deformed state \footnote{Strictly speaking there is also a step length tensor $\mm{l}_0$ in the reference preparation state,  which appears in the neo-classical free energy, in front of $\Lm^{\rm T}$.  However, since the reference state is isotropic, $\mm{l}_0$ is proportional to identity tensor and therefore can be eliminated by redefinition of $\mm{l}$.  }
and describes the statistical conformation of polymer chains in the current state.  It is related to the nematic order parameter $\mm{Q}$ by
\begin{eqnarray}
\mm{l} = a\, \mm{I} - b \, \mm{Q}.
\end{eqnarray}
where $a$ and $b$ are some microscopic constants.
In this work we shall always normalize $\mm{l}$ such that it has determinant one.  In the principle coordinate system of the nematic order parameter, the step length tensor $\mm{l}$ can be represented as a matrix:
\begin{eqnarray}
\mm{l} =
\left(
\begin{array}{ccc}
\frac{1}{\ar}  & 0  & 0  \\
0  & \frac{1}{\ar}  & 0  \\
 0 & 0  &  \ar^2
\end{array}
\right)
= (\ar^2 - \ar^{-1}) \nh \nh + \ar^{-1} \mm{I},
\label{lm-nh}
\end{eqnarray}
where $\ar$ is a monotonic increasing function of the magnitude of the nematic order $S$, whose detailed functional form is irrelevant to our study.  For an achiral nematic elastomer, $\ar$ turns out to be the ratio of spontaneous stretch along the direction of the nematic director when the system enters the nematic phase from the isotropic phase \cite{WarTer96}.  Due to the incompressibility constraint, the system shrinks by factor of $1/\sqrt{\ar}$ in the perpendicular directions.   Finally we note that  in Eq.~(\ref{neo-classical}) a constant term $-3 \mu/2$ is introduced so that the elastic free energy vanishes in the isotropic reference state where $\Lm = \mm{l} = \Im$.

The second part $f_{\Qm}$ in Eq.~(\ref{f-total}) is the Landau-de Gennes free energy for a chiral nematic liquid crystal.   Assuming that the nematic order is well saturated with fixed magnitude $S$ in the cholesteric state, the relevant nematic free energy is the Frank free energy for chiral nematic liquid crystals \cite{LC:deGennes,Cholesteric-Lav-Kle-review}:
\begin{eqnarray}
f_{\rm Frank} &=&  \frac{1}{2} K_1 (\nabla \cdot \nh)^2
 + \frac{1}{2} K_2(\nh \cdot \nabla \times \nh - q_0)^2
 \label{f-Frank-0}\\
 &+&  \frac{1}{2} K_3 (\nh \times \nabla \times \nh)^2
+  K_{24}  \nabla \cdot \left(
\nh \cdot \nabla \nh - \nh \nabla \cdot \nh \right),
\nonumber
\end{eqnarray}
where $K_1,K_2,K_3$, are splay, twist, and bending constants
respectively, while $q_0^{-1} = \ell_0 $ is the cholesteric pitch
for the corresponding chiral nematic liquid crystal.    $K_{24}$
is the saddle splay constant, which plays an important role in the
physics of blue phase \cite{bluephase-1,bluephase-2,LC:deGennes}.
Since the saddle splay density is a complete differential, its
volume integral can be transformed into a surface integral by
Gauss' theorem, and therefore scales the same as the surface
anchoring of the nematic director field, which we shall not
consider in the work.  Nevertheless, it is rather straightforward
to include this surface interaction.  Also, it is important to
note that all the derivatives in Eq.~(\ref{f-Frank-0}),
$\nabla_{i} =
\partial/\partial r_i $, are with respect to the {\em
Eulerian coordinates}, i.e. Cartesian coordinates of mass points
in their deformed states.   This is required by the liquid nature
of Frank free energy: at length scales where the Frank free energy
becomes important, the system is essentially a liquid.  The
physical quantities of a liquid should be naturally expressed in
terms Eulerian coordinates, rather than in terms of Lagrangian
coordinates.   To avoid confusion in notation, we shall use
$\nabla_{i} = \partial/\partial r_i $ for derivative with respect
to Eulerian coordinates and use $\partial_a = \partial /\partial
x_a$ to denote the partial derivative with respect to the
Lagrangian coordinates.

As a first step, let us discuss the total free energy
Eq.~(\ref{f-total}) qualitatively. Within the one constant
approximation of the Frank free energy, and ignoring the surface
saddle splay term for a moment, there are three natural length
scales in this problem. $q_0^{-1} = \ell_0$ is the chirality
pitch, while  $a_0 = \sqrt{ K / \mu}$ is the cross-over length
scale set by the competition between network elasticity and
nematic director elasticity.    The third length scale is the
radius $R$ of the cylinder.  For most liquid crystalline
elastomers, we estimate $K \sim 2 - 4 \times 10^{-12} N$, while
$\mu \sim 10^4 -10^6 Pa$. Therefore $a_0 \sim 1-10 nm$,
constituting the shortest length scale in our problem. On the
other hand, the chirality pitch $\ell_0$ can vary a lot, typically
$0.1 \mu m$ or smaller for strongly chiral materials but may get
much larger for weakly chiral materials.  In particular, it can
even be larger than the cylinder radius $R$.   Also, the regime
$a_0/R \gg 1$ is clearly experimentally inaccessible.


Comparing these three length scales, we are naturally lead to the
following two distinct regimes:
\begin{enumerate}
\item Weak chirality regime $a_0 \ll R \ll \ell_0$.
\item Strong chirality regime  $a_0, \ell_0 \ll R$.
\end{enumerate}
\noindent The strong chirality limit has already been analyzed by
Pelcovits and Meyer \cite{Pelcovits-Meyer-Iso-Cholesteric}, as
well as by Warner \cite{Warner-iso-chiral}.  It is found that as
one tunes the dimensionless ratio $a_0/\ell_0 = a_0 q_0$ to below
a critical value of order of unity, the system goes from a planar
helix director pattern to a conical helix pattern.   In this work,
we shall mainly focus on the weak chirality limit.  Similar to
reference \cite{Pelcovits-Meyer-Iso-Cholesteric} and
\cite{Warner-iso-chiral}, we shall use variational methods,
proposing two kinds of candidate states with certain variational
parameters and minimizing the total free energy over these
parameters.



\section{Double twist of nematic director and twist of cylinder}
Consider a cylindrical block of isotropic chiral elastomer of
radius $R$, aligning along the $z$ axis.   We need to find the
nematic director field $\nh(\xv)$ as well as the elastic
deformation $\rv(\xv)$ relative to the isotropic reference state
that minimizes the total free energy.   One  possible low energy
configuration for the chiral Frank free energy is a double twist
texture, as illustrated in Fig.~\ref{helix}G.  In chiral nematic
liquid crystals, the double twist configuration is energetically
favorable if the saddle splay modulus $K_{24}$ is positive and
large enough \cite{bluephase-1,bluephase-2,LC:deGennes}.
According to the current understanding of the blue phase, these
double twist cylinders pack into a three dimensional periodic
structure with cubic symmetry.   In liquid crystalline elastomers,
due to the nemato-elastic coupling, a double twist nematic
director texture necessarily induces twist of the cylinder,
together with a uniaxial stretch $\lambda$ along the cylinder
axis:
\begin{eqnarray}
\rv(\xv) = \mm{O}_z(\alpha\,z)
\left(
\begin{array}{ccc}
\frac{1}{\sqrt{\lambda}}  & 0  & 0  \\
0  &  \frac{1}{\sqrt{\lambda}} & 0  \\
 0 & 0  &  \lambda
\end{array}
\right) \cdot \xv,
\label{rv-dt}
\end{eqnarray}
where
\begin{eqnarray}
\mm{O}_z (\alpha \,z) =
\left(
\begin{array}{ccc}
\cos \alpha \, z  &  -\sin \alpha\,z  &  0 \\
\sin \alpha\,z  &   \cos \alpha \, z  &  0 \\
 0 & 0  & 1
\end{array}
\right)
\end{eqnarray}
is a rotation about the z-axis by an angle $\alpha\,z$. Using the
cylindrical coordinate system, the Lagrangian coordinates of a
point $\xv$ are given by the triplet $(\rho, \phi, z)$:
\begin{eqnarray}
\xv =
\left(
\begin{array}{ccc}
\rho \, \cos \phi   \\
 \rho \, \sin \phi    \\
z
\end{array}
\right).
\label{xv-ref}
\end{eqnarray}

In the deformed state Eq.~(\ref{rv-dt}), the Eulerian coordinates $\rv(\xv)$ are given by
\begin{eqnarray}
\rv(\xv) =
\left(
\begin{array}{ccc}
r \, \cos \varphi   \\
 r \, \sin \varphi    \\
r_z
\end{array}
\right)
= \left(
\begin{array}{ccc}
\frac{\rho}{\sqrt{\lambda}} \cos(\phi + \alpha \, z)  \\
\frac{\rho}{\sqrt{\lambda}} \sin(\phi + \alpha \, z)    \\
\lambda \,z
\end{array}
\right),
\end{eqnarray}
where we have used  Eq.~(\ref{rv-dt}) and Eq.~(\ref{xv-ref}).
Therefore we find
\begin{eqnarray}
(r, \varphi, r_z )  = (\frac{\rho}{\sqrt{\lambda}}, \phi + \alpha \, z, \lambda\,z).
\label{r-x-dt}
\end{eqnarray}
Note that the deformed cylinder has height $L \lambda$ and radius $R/\sqrt{\lambda}$.

Let $\hat{e}_{\varphi}$ be the unit vector associated with the Eulerian cylindrical coordinate $\varphi$:
\begin{eqnarray}
\hat{e}_{\varphi} = \left| \frac{\partial \rv}{\partial \varphi} \right|^{-1}
\frac{\partial \rv}{\partial \varphi}
=
\left(
\begin{array}{ccc}
- \sin \varphi   \\
 \cos \varphi  \\
 0
\end{array}
\right),
\end{eqnarray}
In terms of the Eulerian coordinates, a double twist texture of nematic director is represented:
\begin{eqnarray}
\nh(\rv) = \hat{e}_{z} \cos\theta(r) + \hat{e}_{\varphi} \sin \theta(r).
\label{nh-dt-1}
\end{eqnarray}
Note that the twist angle $\theta(r) = \theta(\rho/\sqrt{\lambda})$ can be equally well represented as a function of Lagrangian coordinate $\rho$.   $\theta(r)$ satisfies the boundary condition $\theta(0) = 0$, since the nematic director is parallel to $\hat{z}$ on the center axis of the cylinder.   On the outer surface of the cylinder $r = R/\sqrt{\lambda}$, $\theta(R/\sqrt{\lambda})$ is free to vary.

 Calculation of the deformation gradient using Eq.~(\ref{rv-dt}) is a trivial and tedious matter.  On the other hand, by substituting Eq.~(\ref{nh-dt-1}) into Eq.~(\ref{lm-nh}) we can readily calculate the step length tensor $\mm{l}$.  Substituting these results into  Eq.~(\ref{neo-classical}), we find that the spatially dependent elastic free energy density for the proposed double twist solution is given by
\begin{eqnarray}
f_{\rm el} &=& \frac{\mu}{4 \ar ^2 \lambda }
\left(
1+ \lambda ^3 \ar ^3+\alpha ^2 \rho ^2 \ar ^3+3
   \ar ^3+\lambda ^3+\alpha ^2 \rho^2
\right.
\nonumber\\
&-& \left.
   \left(\ar ^3-1\right) \left(\lambda
   ^3-\alpha ^2 \rho ^2-1\right) \cos (2 \theta )
 \right.
\nonumber\\
&-& \left. 2 \alpha  \left(\ar ^3-1\right) \lambda ^{3/2}
   \rho  \sin (2 \theta )
\right).
\label{f-el-dt}
\end{eqnarray}

The spatially dependent Frank free energy density can be calculated by substituting
Eq.~(\ref{nh-dt-1}) into Eq.~(\ref{f-Frank-0}), carefully noting that all derivatives are with respect to the Eulerian coordinates $\rv$.   The result is
\begin{eqnarray}
F_{\rm Frank} &=& \frac{1}{2} K_2\, \left(
\frac{1}{2\,r} \sin 2\theta + \frac{d \theta}{d r}
\right)^2
+ \frac{1}{2} K_3 \, \frac{\sin^4 \theta}{r^2}
\nonumber\\
&-& K_2 q_0\, \left(
\frac{1}{2\,r} \sin 2\theta + \frac{d \theta}{d r}
\right)
- K_{24}\, \frac{\sin2\theta}{r} \frac{d\theta}{dr},
\nonumber\\
\label{f-Frank-dt}
\end{eqnarray}
which is identical to that for a chiral nematic liquid crystal in a double twist cylinder \cite{Cholesteric-Lav-Kle-review}.

In the weak chirality limit, $q_0 R \ll 1$, we expect $\theta(r)$ to be small and linear in $r$.  We can therefore expand the elastic free energy density in terms of $\rho$ \footnote{Remembering that $\rho = \sqrt{\lambda} r$ is proportional to $r$. } and $\theta(r)$:
\begin{eqnarray}
f_{\rm el}  = f_0 + f_2 + \mbox{higher order terms},
\end{eqnarray}
where
\begin{eqnarray}
f_0 &=& \frac{\mu \lambda ^2}{2 \ar
   ^2}+\frac{\ar \mu}{\lambda } - \frac{3}{2} \mu
   , \label{f-0-dt}\\
f_2 &=& \frac{\mu}{2 \ar ^2 \lambda }
\left(
\alpha ^2 \rho ^2 \ar ^3+\left(\ar ^3-1\right)
   \theta ^2 \left(\lambda ^3-1\right)
 \right.
   \nonumber\\
 && \left. -2 \alpha \left(\ar ^3-1\right) \theta  \lambda ^{3/2} \rho
\right)
\label{f-2-dt},
\end{eqnarray}
are terms of order of $r^0$ and $r^2$ respectively.
We shall ignore all higher order terms in the elastic free energy.
Note that $f_0$ is exactly the free energy density for a monodomain nematic elastomer, with anisotropy ratio $\ar$, undergoing a uniaxial deformation coaxial with the step length tensor.  Minimizing $f_0$ over $\lambda$ we obtain
\begin{eqnarray}
\lambda = \ar \longrightarrow f_0 = 0,
\end{eqnarray}
as expected.   Substituting this into Eq.~(\ref{f-2-dt}), we find
\begin{eqnarray}
f_2 \rightarrow \frac{\mu }{2 \ar ^3}
\left(\left(\ar ^3-1\right)
   \theta -\alpha  \ar ^{3/2} \rho \right)^2,
   \label{f-2-2-dt}
\end{eqnarray}
which is a complete square.   Since $f_{\rm Frank}$ is independent of $\alpha$, and since $\theta$ is linear in $\rho$ as will be shown below, Eq.~(\ref{f-2-2-dt}) is minimized by
\begin{eqnarray}
\alpha &=& \ar^{-3/2} (\ar^3 - 1) \frac{\theta}{\rho}, \label{alpha-dt-0}\\
f_2 &=& 0.
\end{eqnarray}
Hence there is {\em no} elastic free energy cost for the double twist state up to the order of $(\alpha R)^2$.  As we shall show below, the parameter $\alpha$ is of order of $q_0$.
Hence  $(\alpha R)^2$ is indeed a small parameter in the weak chirality limit.

Similarly, we expand the Frank free energy in terms of $\theta$
and $r$.  To the leading order we find
\begin{eqnarray}
&&f_{\rm Frank} =
\label{f-Frank-dt-2} \\
&& \frac{1}{2} K_2\, \left(
\frac{\theta}{r}  + \frac{d \theta}{d r}
\right)^2
- K_2 q_0\, \left( \frac{\theta}{r}  + \frac{d \theta}{d r}
\right)
- K_{24}\,\frac{2\,\theta}{r}   \frac{d\theta}{dr},
\nonumber
\end{eqnarray}
which only depends on $\theta$.  Note that the bending term is of higher order when compared to all other terms that we have kept.

We have to minimize the total Frank free energy density
\begin{eqnarray}
F_{\rm Frank} = 2 \pi L \lambda  \int_0^{\frac{R}{\sqrt{\ar}} }
f_{\rm Frank}\, r  dr
\end{eqnarray}
over $\theta(r)$ in order to determine the optimal director
texture.   Let us define $R_{\ar} = R/\sqrt{\ar}$ in order to
streamline the notation below. Calculating the first variation of
the Frank free energy, including the boundary terms at $r =
R_{\ar}$, we find
\begin{eqnarray}
&& \frac{\delta F_{\rm Frank}}{2\pi L \lambda} = K_2 \int_0^{R_{\ar}}
dr \left( - r \,\theta''(r) -\theta'(r) + \frac{\theta(r)}{r}
\right) \delta \theta(r)
\nonumber\\
&& +  \left[ \,K_2 \left( \,\theta(R_{\ar})
+ R_{\ar}  \theta'(R_{\ar})
- R_{\ar} q_0
\,\right)
- 2 K_{24} \theta(R_{\ar})
\,\right] \delta \theta(R_{\ar}).
\nonumber\\
\end{eqnarray}
Since the twist angle $\theta(r)$ is free to vary on the boundary $r = R_{\ar}$, we have to set both the integrand and the boundary term to zero in order to find the minimizing solution. This leads to the following two Euler-Lagrange equations
\begin{eqnarray}
&&r\,\theta''(r) + \theta'(r) + \frac{\theta(r)}{r} = 0,\\
&& \left( \theta(R_{\ar})
+ R_{\ar} \theta'(R_{\ar})
-R_{\ar} q_0
\right)
- 2 \eta \, \theta(R_{\ar}) = 0,
\end{eqnarray}
where $\eta = K_{24}/K_2$ is a dimensionless ratio.  Solving these two equations we find
\begin{eqnarray}
\theta(r) = \frac{q_0}{2 (1-\eta)} \, r
 = \frac{q_0\rho }{2 (1-\eta) \sqrt{\ar}} ,
 \label{theta-rho}
 \end{eqnarray}
which explicitly shows that $\theta(r)$ is indeed linear in $r$.
The twist angle on the  boundary  is given by $$\theta(R_{\ar}) =
\frac{q_0R}{2 (1-\eta) \sqrt{\ar}},$$ which serves as a small
parameter controlling the validity of the perturbative analysis.
Substituting  Eq.~(\ref{theta-rho}) into Eq.~(\ref{alpha-dt-0}) we
find the parameter $\alpha$ given by
\begin{eqnarray}
\alpha = \frac{ (\ar^3 - 1)\,q_0}{2 (1-\eta) \ar^2} ,
\label{alpha-dt}
\end{eqnarray}
which is indeed a constant, of the same order of $q_0$, and
independent of $r$.  Substituting Eq.~(\ref{theta-rho}) into
Eq.~(\ref{f-Frank-dt-2}) we find the Frank free energy density,
which is also the total free energy density (since the elastic
free energy vanishes at the order of $(\alpha R)^2$), to be given
by
\begin{eqnarray}
f_{\rm tot} = f_{\rm Frank} = - \frac{K_2 \eta q_0^2}{2\,(1-\eta)}.
\label{f-dt}
\end{eqnarray}

Summarizing Eq.~(\ref{theta-rho}), Eq.~(\ref{alpha-dt}) and
Eq.~(\ref{f-dt}),  we find that if $\eta <1$, our perturbative
calculation is quantitatively good in the weak chirality regime
where $q_0 R/2(1-\eta) \sqrt{\ar} \ll 1$.  The double twist state
is very efficient in minimizing both the elastic free energy and
the Frank free energy.  In particular, when the saddle splay
constant $K_{24}$ vanishes, $\eta = 0$, and therefore the total
free energy Eq.~(\ref{f-dt}) also vanishes.  Note that the total
free energy is positive definite if $\eta = 0$.  Hence the double
twist state is clearly the ground state, at least up to the order
of $(q_0R)^2$.   By contrast, for a cholesteric liquid crystal
with $K_{24} = 0$, the blue phase is clearly not the lowest energy
state, compared to the usual helical state. This shows that unlike
in the blue phase of cholesteric liquid crystal, the saddle splay
constant $K_{24}$ does not play an important role in the formation
of the double twist pattern in a cholesteric elastomer.  When $q_0
R/2(1-\eta) \sqrt{\ar}$ is comparable or larger than unity, the
higher order terms of the elastic free energy and the Frank free
energy can not be neglected, and one has to minimize the full free
energy Eq.~(\ref{f-el-dt}) and Eq.~(\ref{f-Frank-dt}).   Finally
if $\eta >1$, a perturbative calculation in power of $\alpha R$ is
qualitatively incorrect, no matter how small the parameter $q_0 R$
is.  We must minimize the full elastic free energy
Eq.~(\ref{f-el-dt}) and Eq.~(\ref{f-Frank-dt}).


\section{Helical State}
\label{Sec:helical}

In the weak chirality regime that we are interested in, $q_0 a_0
\ll q_0 R \ll 1$, the elastic energy scale (per unit volume) $\mu$
is much larger than the chiral Frank energy scale $K\,q_0^2$.
Therefore the conical helix state studied in reference
\cite{Pelcovits-Meyer-Iso-Cholesteric} and
\cite{Warner-iso-chiral} can never be the ground state, as it only
partially minimizes both the Frank free energy and the elastic
free energy. There is however, another potential candidate for the
ground state, which can minimize the elastic free energy up to the
leading order.  Let us consider a planar helix director pattern
along the cylinder axis, where the nematic director remains
perpendicular to the cylinder  z-axis and rotates around this axis
with pitch $\alpha$ \footnote{We note that $\alpha$ is the helical
pitch measured by the Lagrangian coordinate $\xv$.  The physical
value of the pitch however, should be defined using the Eulerian
coordinate and is therefore given by $\alpha\sqrt{\ar}$. }:
\begin{eqnarray}
\nh(z) =    \eh_x \, \cos \alpha \,z
    +    \eh_y \, \sin \alpha \, z
=  \mm{O}_z(\alpha \,z) \, \eh_x .
    \label{Q-ansatz-1}
\end{eqnarray}
In the following, we shall use both dyadic notation and matrix
notation of tensor quantities. The corresponding local step length
tensor is given by
\begin{eqnarray}
\mm{l}(z) = \mm{O}_z(\alpha\,z)  \mm{l}(z = 0)
\mm{O}_z(- \alpha\,z),
\end{eqnarray}
where
\begin{eqnarray}
\mm{l}(z=0) =
\left(
\begin{array}{ccc}
 \ar ^2  & 0 & 0 \\
 0 & \frac{1}{\ar } & 0 \\
0 & 0 & \frac{1}{\ar }
\end{array}
\right)
\end{eqnarray}
is the step length tensor at the plane $z = 0$.   This variational form of nematic director field is the same as the planar helix sate considered in reference \cite{Pelcovits-Meyer-Iso-Cholesteric}.

Due to the nemato-elastic coupling, the polymer network prefers to
stretch along the local nematic director.   This however implies
that the direction of local strain deformation rotates by an angle
$\pi/2$ between two cross sections $\ell_0/4$ apart along the
cylinder. This leads to an additional strain energy density
$\mu\,(\alpha R)^2$ that is quadratic in the cylinder radius. For
a fat cylinder (or in the strong chirality limit), $\alpha R \gg
1$ and this strain energy is prohibitively high \footnote{In the
fat cylinder/strong chirality limit, the system would prefer a
uniform uniaxial deformation along the cylinder axis instead, as
studied by Pelcotivs and Meyer, as well as by Warner. }.   For a
thin cylinder (or in the weak chirality limit), however, $\alpha R
\ll 1$ and this additional strain energy only constitutes a
perturbation to the strain energy of the corresponding uniform
deformation.  Nevertheless, to reduce the additional strain energy
at the order of $\mu\,(\alpha R)^2$, the system can globally twist
in the direction opposite to the nematic helix.  The overall
nonuniform deformation, shown in Fig.~\ref{helix}B, is represented
by the Eulerian coordinates as functions  of the Lagrangian
coordinates:
\begin{eqnarray}
&&\rv(\xv) \equiv \tilde{\Lm}(z) \cdot \xv
 \label{deformation-ansatz-1} \\
 &=&
\mm{O}_z(\alpha\,z)
\left(
\begin{array}{ccc}
\lambda & 0 & 0  \\
 0 & \frac{1}{\sqrt{\lambda}}& 0 \\
 0 & 0 &  \frac{1}{\sqrt{\lambda}}
\end{array}
\right)
\mm{O}_z(-(\alpha+\beta)\,z)
 \cdot \xv,
\nonumber
\end{eqnarray}
where $\beta$ measures the global twist of the solid. We also note
that due to the additional inhomogeneous strain deformation, the
inverse pitch of the director helix $\alpha$ is generically
different from the inverse pitch $q_0$ of the corresponding liquid
crystal  system.   This will be made clear in the calculation
below.

It is important to note that the matrix $\tilde{\Lm}$ defined through Eq.~(\ref{deformation-ansatz-1}) explicitly depends on the coordinate $z$ and is {\em not} the deformation gradient $\Lambda_{ia}$.   The latter should be obtained by taking partial derivative of Eq.~(\ref{deformation-ansatz-1}) with respect to Lagrangian coordinates $x,y,z$.  Being derived in this way, the deformation gradient matrix naturally satisfies the following compatibility conditions:
\begin{eqnarray}
\partial_a \Lambda_{ib} = \partial_b \Lambda_{ia}. \nonumber
\end{eqnarray}

Substituting the deformation gradient and nematic order parameter Eq.~(\ref{Q-ansatz-1}) into Eq.~(\ref{neo-classical}), integrating over the reference volume of the cylinder, and dividing it by the total volume $\pi R^2 L$,
we obtain the elastic free energy density for the proposed deformation gradient as
\begin{eqnarray}
&& f_{\rm el}[\alpha, \beta, \lambda] = f_0 + f_2 ,
\label{f-elastic}\\
&& f_0 =  \mu \left(
\frac{\ar}{\lambda} + \frac{\lambda^2}{2\,\ar^2}
\right) - \frac{3}{2} \mu,
\label{f-0}
\\
&& f_ 2 =
 \frac{\mu R^2 }{8 \ar ^2 \lambda } \left[
\left(\ar ^3+\lambda ^3\right) \beta ^2
-2  \left(\ar ^3-\lambda ^{3/2}\right)
\left(\lambda^{3/2}-1\right)  \alpha \beta
\right. \nonumber\\
&&  \left.\,\,\,   +\alpha ^2 \left(\ar^3+1\right)
\left(\lambda ^{3/2}-1\right)^2 \right]
.\label{f-2-1}
   \end{eqnarray}
Note that $f_0$ is independent of the cylinder radius $R$ and is
identical to the free energy of an achiral nematic elastomer
undergoing uniaxial and homogeneous deformation.    By contrast,
$f_2$ is proportional to $R^2$, and quadratic in $\alpha$ and
$\beta$.  $f_2$ is clearly due to the inhomogeneous deformation.
The ratio between $f_0$ and $f_2$  scales as $(\alpha R)^2$ as
discussed earlier.  The dimensionless ratio $\alpha R$
characterizes the importance of chirality in this problem.  In
reference \cite{Pelcovits-Meyer-Iso-Cholesteric} and
\cite{Warner-iso-chiral} this ratio is implicitly taken to be
large at the very beginning.  In this work, we shall assume it to
be a small number.  More precisely we shall assume another
dimensionless ratio $q_0\,R \ll1$.  Also we shall see below that
for this proposed variational solution, the Frank free energy
scales the same as $f_2$,  hence it is reasonable to first
minimize $f_0$ and then the sum of $f_2$ and $f_{\rm Frank}$.
Minimization of $f_0$ leads to
\begin{eqnarray}
\lambda &=& \ar,\\
f_0 &\rightarrow&  0.
\end{eqnarray}
That is, the local elastic deformation is identical to that of a homogeneous achiral nematic elastomer.  Inclusion of $f_2$ and $f_{\rm Frank}$ induces small change of $\lambda$ at order of $\alpha R$.

The Frank free energy density for the proposed director pattern
Eq.~(\ref{Q-ansatz-1}) can also be easily calculated. Again we
have to be careful with the derivatives in Eq.~(\ref{f-Frank-0})
that are with respect to the Eulerian coordinate $\rv$.   After
some tedious calculation and replacing $\lambda$ with $\ar$, we
find
\begin{eqnarray}
f_{\rm Frank} = \frac{1}{2}K_2 \alpha ^2 \ar
- K_2 \alpha q_0  \sqrt{\ar}
+\frac{1}{2} K_2 q_0^2,
   \label{f-Frank}
\end{eqnarray}
which is independent of elastic constants $K_1$, $K_3$ and $K_{24}$.

We still need to minimize the sum of $f_2$, given in  Eq.~(\ref{f-elastic}) and $f_{\rm Frank}$ Eq.~(\ref{f-Frank}), over the remaining two variational parameters $\alpha, \beta$.  Since $f_{\rm Frank}$ does not depend on $\beta$, we minimizes $f_2$ over $\beta$ and find
\begin{eqnarray}
\beta &=& \frac{\alpha  \left(\ar ^{3/2}-1\right)^2}{2 \ar^{3/2}}
  , \label{beta}\\
f_2 &=& \frac{R^2 \alpha ^2 \left(\ar ^3-1\right)^2
   \mu}{16 \ar ^3}
   . \label{f-2-2}
\end{eqnarray}
We note that as long as $\ar \neq 1$, and $\alpha \neq 0$, the global spontaneous twist of solid $\beta$ does not vanish.  More importantly, unlike the double twist state, the planar helix state considered here does cost elastic free energy at the order of $(\alpha R)^2$.

We can now minimize the sum of Eq.~(\ref{f-2-2}) and Eq.~(\ref{f-Frank}) over $\alpha$,
which leads to
\begin{eqnarray}
  \alpha =  \frac{8 a_0^2q_0 \ar ^{7/2}}{8 a_0^2
   \ar ^4+R^2 \left(\ar ^3-1\right)^2}
\end{eqnarray}
Remembering $a_0/R \leq 10^{-6}$ even for $R = 1mm$, and $\ar \neq 1$, the first term in the denominator can be safely ignored and $\alpha$ can be approximated as
\begin{eqnarray}
\alpha =\frac{8 a_0^2 q_0 \ar ^{7/2}}{R^2
   \left(\ar ^3-1\right)^2}
  \sim  \left(\frac{a_0}{R}\right)^2 \,q_0
    \ll q_0.
      \label{alpha-scaling}
\end{eqnarray}
This result indicates that the helix of nematic director is strongly resisted by the nemato-elastic coupling energy and the pitch becomes much longer than the corresponding value $\ell_0$ in the nematic liquid crystals.
Substituting Eq.~(\ref{alpha-scaling}) into Eq.~(\ref{f-2-2}) and Eq.~(\ref{f-Frank}), we find
\begin{eqnarray}
f_{\rm tot}  \approx
- \frac{a_0^2}{R^2}
\frac{4 K_2 q_0^2 \ar ^4}
{\left(\ar ^3-1\right)^2 }
+ \frac{1}{2} K_2 q_0^2
\approx
 \frac{1}{2} K_2 q_0^2.
  \label{f2-ffrank}
\end{eqnarray}
The total free energy is therefore {\em positive}, in strong contrast with the double twist state we considered in the preceding section.  The planar helix state considered here is therefore not efficient in energy minimization.   This is clearly due to the extra elastic free energy cost Eq.~(\ref{f-2-2}) caused by the nemato-elastic coupling.

\vspace{5mm}

\section{Discussion and Conclusion}
We have shown in this work that in the weak chirality limit $q_0 a_0 \ll q_0 R \ll 1$, the double twist state minimizes both the Frank free energy and the elastic free energy up to the order of $(q_0 R)^2$, and is therefore a good candidate for the real ground state.   The planar helix state, on the other hand, is strongly influenced by the nemato-elastic coupling, with the pitch much longer than the corresponding value in cholesteric liquid crystal.
As the dimensionless parameter $q_0 R$ becomes larger than one, the elastic energy cost due to inhomogeneous strain, scaling as $\mu (q_0 R)^2$,
dominates all other terms.   When $q_0 R \gg 1$ and $q_0 a_0 \gg 1$, the ground state is likely to be the conical helix state with $\theta \approx \pi/2$, according to th	e studies in references \cite{Pelcovits-Meyer-Iso-Cholesteric,Warner-iso-chiral}.   The conical state and the double twist state are qualitatively different, and can not be mutually accessed in a continuous fashion.  Therefore the aforementioned two regimes must be separated by a first order phase transition, located around $q_0 R \sim 1$.   Study of this transition is technically challenging and is beyond the scope of this work.

We acknowledge financial support from the American Chemical Society under grant PRF 44689-G7.

\bibliography{reference-all}
\end{document}